\documentclass[aps,reprint,prl,superscriptaddress,showpacs,floats]{revtex4-1}
\usepackage{graphicx}
\usepackage{amsmath,graphicx,dcolumn}
\usepackage{hyperref}
\usepackage[usenames]{color}
\usepackage{datetime}
\usepackage{color}
\usepackage{textcomp}
\usepackage{ulem}

\newcommand*{\ur}{URu$_2$Si$_2$}

\newcommand*{\an}{{{\AA}$^{-1}$}}

\begin{document}

\title{Photoemission Imaging of 3D Fermi Surface Pairing at the Hidden Order Transition in {\ur} }

\author{Jian-Qiao Meng}
\affiliation{Condensed Matter and Magnet Science Group, Los Alamos
National Laboratory, Los Alamos, NM 87545, USA}

\author{Peter M. Oppeneer}
\affiliation{Department of Physics and Astronomy, Uppsala University, Box 516, S-75120 Uppsala, Sweden }

\author{John A. Mydosh}
\affiliation{Kamerlingh Onnes Laboratory, Leiden University, NL-2300 RA Leiden, The Netherlands }

\author{Peter S. Riseborough}
\affiliation{Department of Physics, Temple University - Philadelphia, PA 19122, USA}

\author{Krzysztof Gofryk}
\affiliation{Condensed Matter and Magnet Science Group, Los Alamos
National Laboratory, Los Alamos, NM 87545, USA}

\author{John J. Joyce}
\affiliation{Condensed Matter and Magnet Science Group, Los Alamos
National Laboratory, Los Alamos, NM 87545, USA}

\author{Eric D. Bauer}
\affiliation{Condensed Matter and Magnet Science Group, Los Alamos
National Laboratory, Los Alamos, NM 87545, USA}

\author{Yinwan Li}
\affiliation{Wolfram Research Inc., Champaign, IL 61820, USA}

\author{Tomasz Durakiewicz \bigskip}
\email[]{corresponding author: tomasz@lanl.gov}
\affiliation{Condensed Matter and Magnet Science Group, Los Alamos National Laboratory, Los Alamos, NM 87545, USA}

\date{\today}

\begin{abstract}
We report angle-resolved photoemission spectroscopy (ARPES) experiments probing deep into the hidden order (HO) state of {\ur}, utilizing tunable photon energies 
with sufficient energy and momentum resolution to detect the near Fermi surface (FS) behavior. 
Our results reveal: (i) the full itinerancy of the 5$f$ electrons; (ii) the crucial three-dimensional (3D) $k$-space nature of the FS and its critical nesting vectors, in good comparison with density-functional theory calculations, and (iii) the existence of hot-spot lines and pairing of states at the FS, leading to FS gapping in the HO phase.

\end{abstract}
\maketitle

The physics underlying the Hidden Order (HO) transition in {\ur} remains elusive after over 25 years of intensive research \cite{Palstra85, Schlabitz86, Maple86, MydoshRMP11}. This second-order phase transition at $T_0$= 17.5 K is marked by a jump in specific heat and removal of significant fraction of total entropy \cite{Palstra85, Maple86}, however, with no sign of static dipolar magnetic order except under pressure \cite{Amitsuka07, Villaume08}. The appearance of a new 5$f$ electronic phase with apparent lack of ordered magnetism has given birth to the mystery of the unknown, or `hidden' order parameter, addressed so far by a few dozens of theories (e.g., \cite{Kasuya97, Santini98, Chandra02, Varma06, Elgazzar09, Balatsky09, Haule09, Harima10, Schmidt10, Dubi11, MydoshRMP11}), and has recently been proposed to occur in a wide range of materials \cite{Riseborough13}. 

One of the important dividing lines between different HO theories is related to time-reversal symmetry breaking \cite{Chandra02, Ikeda98, Ikeda12} as opposed to time-reversal invariance \cite{BarzykinPRL93, Kasuya97, Haule09, Harima10}. Either commensurate \cite{Tripathi07, Elgazzar09, Kasuya97, Ikeda98, Harima10} or incommensurate \cite{Chandra02, Balatsky09, Dubi11, Su11, TanmoyDas12} FS renormalization was suggested as a driver of the HO transition, with a magnetic resonance mode \cite{Elgazzar09,Oppeneer10},  hybridization wave \cite{Balatsky09, Dubi11}, or orbital-density wave \cite{Riseborough12} being responsible for the gap formation. Another dichotomy in our understanding of {\ur} is related to the $f$ occupancy, with multipolar ordering often invoked \cite{Haule09, Harima10, Hanzawa07, Santini98, Chandra02, Cricchio09, Kotliar11}, where the mainly localized $5f^2$ configuration is in opposition to the itinerant 5$f$ configuration and FS instability models \cite{Varma06, Elgazzar09, Balatsky09, Oppeneer11, Haraldsen11, Dubi11, TanmoyDas12, Riseborough12}. Recent 
comparisons of theory and experiment \cite{Oppeneer10, MydoshRMP11,Chandra13} reveal that the localized--itinerant division in theoretical approaches is still very much alive. Several recent experimental efforts \cite{Okazaki11, Walker11, Bourdarot11, MKLiu11, Aynajian10, Santander09, Yokoya10, Yokoya12, Dakovski11} have revealed aspects of the HO, however, further critical experiments are needed to distinguish between the wide m{\'e}lange of theories.

With the FS instability emerging as a likely mechanism for HO, the electronic structure and FS topology of {\ur} both become crucial in understanding the HO state. Modern ARPES techniques provide a detailed view of the hybridized bands of a strongly correlated system, in addition to a full momentum determination of the 3D FS throughout the Brillouin zone (BZ) \cite{Damascelli03}.

Since its discovery,  {\ur} has been the subject of photoemission investigations \cite{Allen87, Santander09, Yokoya10, Yokoya12, Denlinger01, Kawasaki11}. Initially, different ARPES experiments produced apparently contradictory results: a heavy band collapsing towards the Fermi level though the transition \cite{Santander09} or a heavy band developing below the Fermi level \cite{Yokoya10} or specific hot spots forming on parts of the Fermi surface \cite{Dakovski11}, with a different value of the Fermi momentum ($k_F$) in each experiment. Definitely, the 3D nature of the FS needs to be taken into account by extending ARPES beyond one photon energy by utilizing variable photon energies to probe the electronic structure in {\ur} as a function of $\boldsymbol{k}$. Recent investigations describe the initial attempts in reconciling these issues \cite{Boariu13}. However, the full 3D topology of the FS of {\ur} and its reconstruction at $T_0$ have not yet been revealed.

In this Letter we report ARPES results, using tunable synchrotron radiation, on high-quality single crystals of {\ur}. A comparison of spectra dominated by $d$-band spectral weight with spectra with U $5f$ weight resonantly enhanced, establishes the basic itinerant properties of the $5f$ electrons. 
Next, by utilizing detailed normal emission ARPES, we map the Fermi surface of {\ur} along the $k_z$ (perpendicular) momentum direction in high resolution, identifying nesting vectors and those parts of the FS that are renormalized upon HO transition. The measured FS topology is compared with density-functional theory (DFT) calculations \cite{Elgazzar09, Oppeneer11}. Constant photon energy ($h\nu$=51 eV) mapping clearly reveals the existence of hot-spot lines,  due to FS nesting,  along multiple scattering vectors that pinpoint the FS  removal caused by `pairing' of FS states  in the HO phase. 

High resolution angle-resolved photoemission measurements were carried out at PGM beamline 71A of Synchrotron Radiation Center,  
using a SCES4000 hemispherical electron energy analyzer.  All samples were cleaved \textit{in situ} and measured in ultra-high vacuum with base pressure better than 4$\times$10$^{-11}$ mbar. Off- and on-resonance spectra ($h\nu$=92 eV and 98 eV, respectively) were measured at a low sample temperature of 10\,K with total energy resolution $\sim$20\,meV. Wide angle Fermi-edge band ($k_x$-$k_z$ plane) maps were measured in the photon energy range 14\,eV to 36\,eV in steps of 0.25\,eV, with varying energy resolution (15--18 meV) and at a low temperature of 12\,K.  At sample temperatures down to 8\,K, a photon energy of 
51\,eV with energy resolution of $\sim$20\,meV was chosen to probe a Fermi surface ($k_x$-$k_y$ plane) where $k_z$ is close to the $\Gamma$ point. Complementary temperature dependent high-resolution spectra were obtained with a photon energy of 98\,eV with an energy resolution of $\sim$25\,meV. For all these measurements, the angular resolution was 0.2$^{\circ}$. High quality single crystals of {\ur} were grown via the Czochralski technique in electrical tri-arc and tetra-arc furnaces, followed by a 900$^{\circ}$C anneal in Ar for 1 week in the presence of a Zr getter.

\begin{figure}[tbp]
\vspace*{-0.2cm}
\begin{center}
\includegraphics[width=1\columnwidth,angle=0]{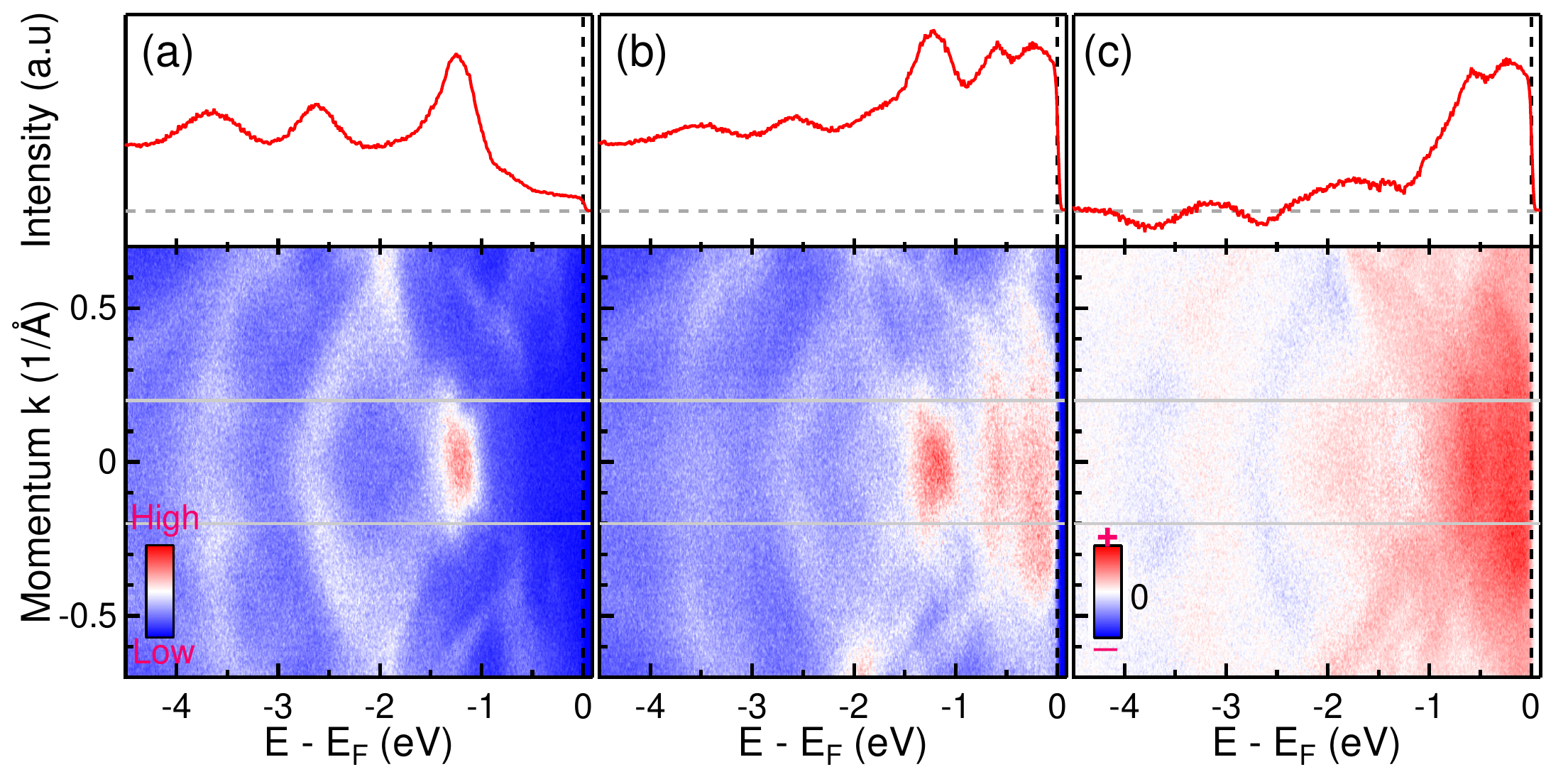}
\end{center}
\vspace*{-0.7cm}
\caption{(color online) \textbf{a\,-\,b}, Off- and on-resonance, respectively, valence band structure near the $\Gamma$-point of {\ur} at 10\,K with 92\,eV ($k_z$ slightly below $\Gamma$) and 98\,eV ($k_z$ slightly above $\Gamma$). \textbf{c}, The intensity difference between on- and off-resonance valence band structure. Blue color depicts decrease of spectral weight, while red denotes increased spectral weight. Top panels of \textbf{a\,-\,c} show the integrated energy distribution curves (EDCs) between the two gray lines.}
\vspace*{-0.4cm}
\end{figure}

Figures 1a and 1b present off- and on-resonance valence spectra (U $5d$$\rightarrow$$5f$), respectively, along the (0,0,1) direction near the $\Gamma$ point at a sample temperature of 10\,K. 
Close to the Fermi energy ($E_F$), the off-resonance spectrum ($h\nu$=92\,eV) shows a density of states of non-$f$ orbital character. The on-resonance spectrum ($h\nu$=98 eV), with intensity of the $5f$ states enhanced via the $5d$--$5f$ resonance, shows a clear gathering of the U $5f$ spectra weight at the Fermi level. The two energies were chosen to allow both the resonance and off-resonance conditions and also to probe the FS on both sides of the $\Gamma$ point. This permits us to neglect the variations due to the 3D nature of the FS and extract only the orbital character from difference spectra.
The difference plot between off- and on-resonance spectra is shown in Fig.\ 1c and reveals almost only pure $5f$ character.
The red color near $E_F$ indicates the density of dispersive and hybridized $5f$ bands participating in bonding and forming the FS. Such a picture is contrary to a localized configuration, e.g.\ $5f^2$, where the majority of the $5f$-weight would be shifted well below and away from the Fermi level. This evidence demonstrates 
that the itinerant U $5f$-electron model \cite{Chandra02, Varma06, Elgazzar09, Balatsky09, Oppeneer11, Haraldsen11, Dubi11, Riseborough12, TanmoyDas12} is an appropriate starting point for describing the HO transition. 

\begin{figure}[tbp]
\vspace*{-0.2cm}
\begin{center}
 \includegraphics[width=1\columnwidth,angle=0]{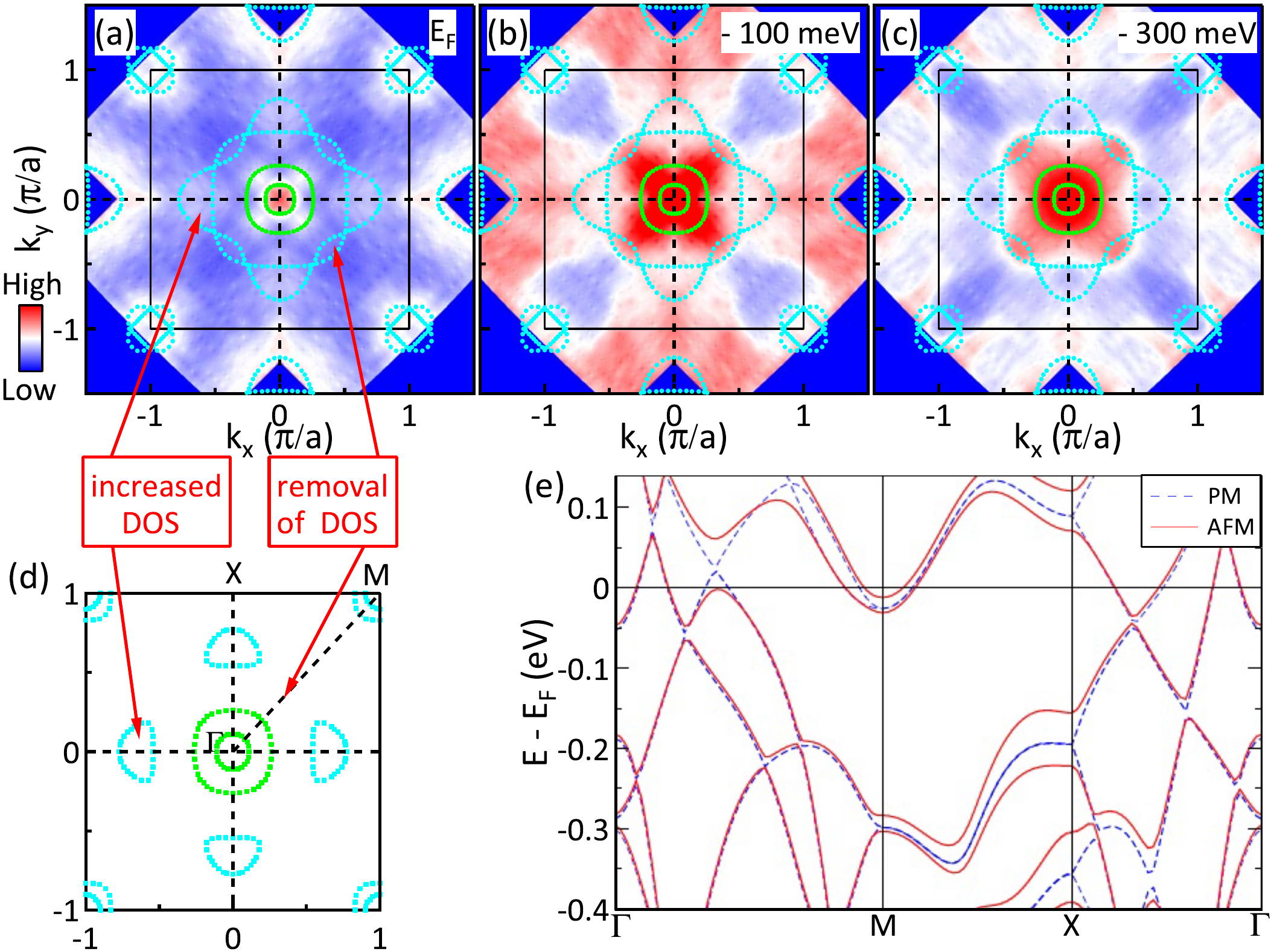}
\end{center}
\vspace*{-0.5cm}
\caption{(color online) Fermi surface mapping in the $k_x$-$k_y$ plane ($k_z$=0) of {\ur} in the HO state. \textbf{a - c}, Constant-energy contours of the band structure at the Fermi level, and at 100\,meV and 300\,meV below Fermi level. The calculated cross sections (light blue circles) of the paramagnetic (PM) FS in the $k_z$=0 plane of the simple tetragonal BZ serve as a guide for the eyes. \textbf{d}, Cross section of the antiferromagnetic (AFM) Fermi surface in the $k_z$=0 plane. Adapted from Ref.\ \cite{Oppeneer10}. \textbf{e}, DFT band dispersions of  {\ur} in PM and AFM phases plotted in the st BZ. Adapted from Ref.\ \cite{Oppeneer11}.}
\vspace*{-0.4cm}
\end{figure}

Figure 2 shows the $k_x$-$k_y$ cut though the FS measured in the HO state at 12\,K using a constant photon energy of 51\,eV, which corresponds to a cut at the $\Gamma$ point. To facilitate comparison with DFT calculations we have overlaid the experimental FS mapping with calculated FS contours \cite{Elgazzar09, Oppeneer10}. The {\it ab initio} computed FS cross-sections are those of paramagnetic (PM) {\ur}, folded over the commensurate nesting vector, 
$\boldsymbol{Q}_0$=$(0,0,1)$, i.e., folded from the bct to the simple tetragonal (st) BZ. The reason for choosing this representation is that it provides maximum clarity on the HO gapping. Figure 2a compares the measured and computed FS cross-sections at $E_F$. At the M-point of the st BZ the presence of a ``double" pocket can be clearly seen, in agreement with the band-structure calculations that predict here two electron pockets. The fact that there is a double pocket verifies that in the HO phase the FS is folded over $\boldsymbol{Q}_0$, i.e., bct lattice periodicity is broken, as it has been predicted earlier \cite{Elgazzar09} and  was deduced from quantum oscillation experiments \cite{Hassinger10}. At the $\Gamma$-point there is first a small pocket, which can be seen in both experiment and calculations.  Next, around $\Gamma$ a second but weakly visible larger pocket appears. Note that this FS sheet is actually folded from the Z-point in the bct phase to $\Gamma$ in the st phase; its presence will become more clear below. 
The next larger FS sheets consist of a quatrefoil entangled with a rounded rectangle. At the intersection of the these two sheets an intriguing FS reconstruction happens. Four half-ellipsoids in $k_x$, $k_y$ directions are present (these can be best seen at the outer sides of the plot). However, in the  $(1,1,0)$ $\Sigma$ directions between the half-ellipsoids there is no intensity and hence no FS cross-section present. This observation is consistent with the prediction that it is the parts between the half-ellipsoids that are gapped in the HO  \cite{Elgazzar09, Oppeneer10}. 
Thus, it can be concluded that the measured and computed FS cross-sections are in good agreement. We ought to mention, first, that there is a narrow intensity going streak-wise outward from $\Gamma$ in the four equivalent $(1,1,0)$ directions, whose origin is presently not known. Second, a four-fold symmetry breaking was recently deduced from torque measurements on tiny crystals, but for our large crystals this effect is not expected to be detectable \cite{Okazaki11}.

Next,  we lower the binding energy to study how the FS contours change for $E$$<$$E_F$; the resulting maps are given in Figs. 2b and 2c. The ``double" pockets at the M-point disappear after $E$ being lowered by 100\,meV, indicating that these are shallow electron pockets. The small pocket at $\Gamma$ becomes less clearly discernible, but it doesn't disappear; according to the band-structure it should disappear at -100 meV. The larger $\Gamma$-centered sheet becomes more expanded at -300\,meV; the half-ellipsoids become weaker and disappear. The most salient change is the appearance of new intensity, already at -100\,meV, extending from $\Gamma$ in the four $\Sigma$ ($\Gamma$--M) directions, in the areas where no intensity was present at $E_F$. This intensity can be due to a renormalized band lying below $E_F$ in the FS gapped area, see Fig.\ 2e. These results suggest that there is, at $E_F$, no FS pocket along $\Sigma$ in the HO, but for lower energy pockets do appear at these loci. We remark, first, that a recent investigation proposed the presence of a cage-$\Sigma$ FS at these positions \cite{Ikeda12} at the Fermi energy, but this is not supported by our measurements. Second, as the lifting of a degenerate band crossing along $\Gamma$-M leads to a renormalized band that moves below $E_F$ (Fig.\ 2e) a FS gapping is expected in this $k$-area. The down-shift of the band, related to the HO gapping, is not too small; our energy resolution is sufficient to detect it. This indicates that the gap formed over the FS in the HO phase is strongly anisotropic. 
 
\begin{figure}[tp]
\begin{center}
\includegraphics[width=1\columnwidth,height=3.3cm, angle=0]{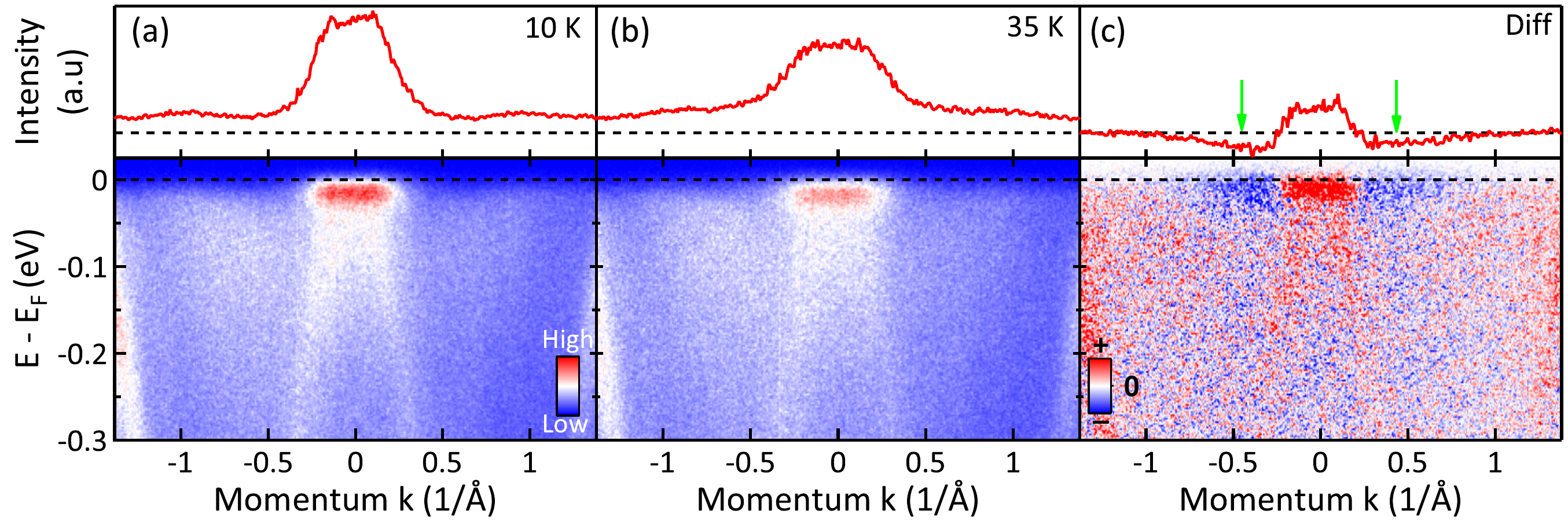}
\end{center}
\vspace*{-0.5cm}
\caption{(color online) Temperature evolution of momentum distribution curves (MDCs) of {\ur} measured near $E_F$ along the (1,1,0) direction, taken with 98\,eV photons at the $5d$--$5f$ resonance and at the $\Gamma$ point. Corresponding ARPES maps from which MDCs at $E_F$  were extracted are shown in each panel, with high intensity areas corresponding to 5f spectral density strongly enhanced by resonance. \textbf{a}, MDCs at 10\,K in the HO;  \textbf{b}, MDCs in the normal state at 35\,K;  \textbf{c}, Difference spectrum showing the disappearance of intensity in HO state at certain parts along the (1,1,0) direction, marked with green arrows.  }
\vspace*{-0.4cm}
\end{figure}
To examine if the sizable FS renormalization predicted for the HO phase along (1,1,0) in the $k_z$=0 plane \cite{Elgazzar09} indeed occurs, we show in Fig.\ 3  the temperature dependence of momentum distribution curves (MDCs) measured with $5f$-sensitive $h\nu$=98 eV. The location in $k$-space is the same as for Fig.\ 1b, only with broader momentum range and slightly worse energy resolution, allowing for temperature dependence study. 
The difference spectrum of MDCs measured in the HO (at 10\,K) and in the normal state (at 35\,K) is revealing.  Evidently in the HO state,  density is removed near $k$=0.3$\frac{\pi}{a}$. This location corresponds to the loci of the two larger, intersecting  FS sheets in Fig.\ 2b, where we noted low density regions. Since Fig.\ 3b shows that these regions exhibit higher densities in the normal state, we can conclude that at these locations 5$f$ density is indeed being removed through FS gapping at the HO transition. 

\begin{figure*}[tbp]
\begin{center}
\includegraphics[width=2\columnwidth,height=4.8cm,angle=0]{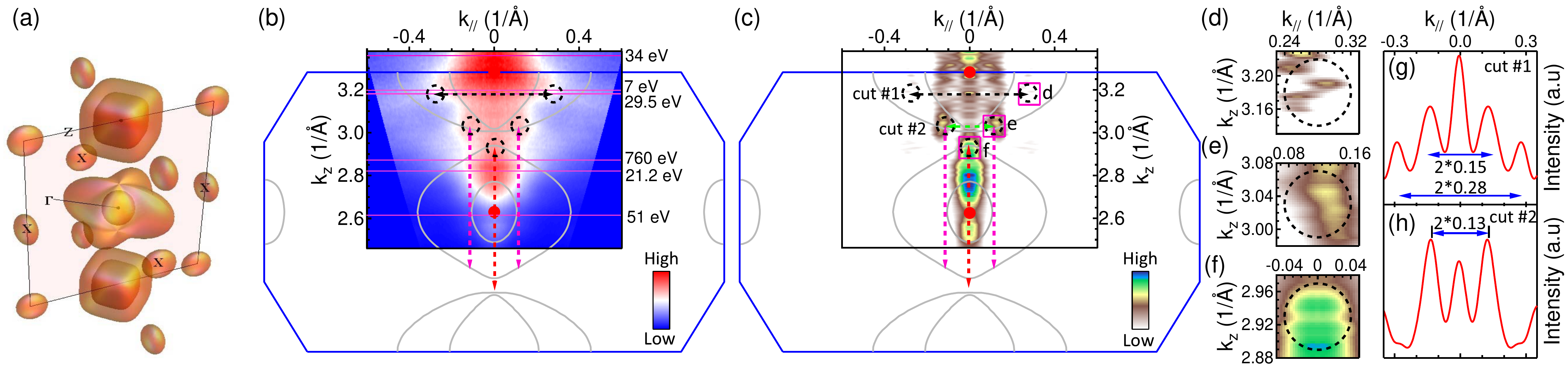}
\end{center}
\vspace*{-0.7cm}
\caption{(color online) Normal emission ARPES study of {\ur} in the HO state. \textbf{a}, Schematic model of the 3D FS in the PM phase. \textbf{b}, Experimental 3D FS map of the Fermi level intensity constructed out of 88 $E_F$ cuts along $k_z$ measured with $h\nu$=14 - 34 eV photons in 250\,meV steps,  in the $\Gamma$-X-Z plane indicated by black lines in \textbf{a}. DFT calculated Fermi contours (solid gray lines) \cite{Oppeneer10} and final state arcs for different photon energies (thin magenta lines) are indicated. The dashed-magenta lines and dashed-black circles present the nesting vectors and hot-spots, respectively. 
\textbf{c}, Second derivation plot of the normal emission given in \textbf{b}. [\textbf{d-f}], Enlarged views of the hot-spot indicated in  \textbf{b}. [\textbf{g-h}], Intensity cuts corresponding to the dashed-black and dashed-green horizontal lines in \textbf{c}. }
\vspace*{-0.4cm}
\end{figure*}

We now investigate the `pairing' interaction between states at the Fermi edge, i.e.,  quasiparticle (QP) mediated coupling of FS states. Figure 4b shows measured Fermi-edge intensity obtained from photon energy dependent normal emission in the HO state at 12\,K. The measurements were performed in a section of the high symmetry plane spanned by $k_z$ and $(1,1,0)$;  the corresponding $k_z$  range covers  more than half a BZ and includes the bct $\Gamma$ and Z points. For experimental convenience, we have selected the photon energy interval from 14 eV to 36 eV;  the energy resolution varies with $h \nu$ between 15 and 18\,meV. The second derivative of raw data is shown in Fig.\ 4c. For comparison we have overlaid the measured intensities with DFT calculated Fermi contours of PM {\ur} in the bct phase (gray curves) \cite{Oppeneer11}.  A schematic drawing of the 3D FS in the PM phase is shown in Fig.\ 4a.

\begin{table}[b]
\vspace*{-0.7cm}
\caption{Selection of Fermi vectors obtained from various measurements corresponding to different values of $k_z$.}
\scriptsize\rm
\begin{tabular}{ccccc}\hline
Method & $h\nu$ (eV) & Direction & Length & Reference\\\hline
ARPES & 21.2 &  $(1,0,0)$  &  $\sim$0.15 ({\an})  &  \cite{Santander09} \\[0.5ex]
ARPES & 21.2 & $(1,1,0)$  &  $\sim$0.20 ({\an})  &  \cite{Santander09} \\[0.5ex]
ARPES & 7.0 & $(1,0,0)$  &  $\sim$0.12 ({\an})  & \cite{Yokoya10} \\[0.5ex]
ARPES & 7.0 & $(1,1,0)$  &  $\sim$0.15 ({\an})  & \cite{Yokoya10} \\[0.5ex]
Tr-ARPES & 29.5 & $(1,1,0$)  &  0.28$\pm$0.04 ({\an})  & \cite{Dakovski11} \\[0.5ex]
STM & $\relbar$ & $(1,0,0)$  &  $\sim$0.15 (2$\pi /a$)  & \cite{Schmidt10} \\[0.5ex]
\hline
\end{tabular}
\end{table}

The second derivative maps are prepared and shown in Fig. 4c to enhance the small variations in intensity corresponding to momentum-dependent QP states.  Cut \#1 shown in Fig.\ 4g and a hot spot in Fig.\ 4d correspond to a cut through the area below the Z point where both Fermi surfaces centered around the Z point are indicated by hotspots, matching Fermi vectors of 0.15 {\an} and 0.28 {\an}. Now reconciled, these vectors were previously seen separately with 7eV  \cite{Yokoya10} and with 29.5eV \cite{Dakovski11} photons, respectively.   
In Figs.\ 4b, c and d the strongest, locally enhanced QP density can be seen on the outer Fermi sheet centered at the bct Z point.  These QP densities 
fit the nesting vector $\boldsymbol{Q}_0$=(0,0,1) connecting the larger, Z-centered squarish FS sheet and the large $\Gamma$-centered quatrefoil (Fig.\ 4a).
This nesting vector is marked by the vertical dashed magenta arrows in Figs.\ 4b and 4c. The enhanced fragments corresponding  to hot-spot areas of increased QP density and FS gapping are shown in Figs.\ 4e and 4f.  A cut (\#2) through the FS including 4e location reveals a small Fermi vector of 0.13 {\an} given in Fig.\ 4h. 
Note that these regions of enhanced QP intensity (Fig.\ 4c) are not point-like, but rather resemble the shape of two lines stretched vertically in the $k_z$ direction along the contour of the Fermi surface. However, the areas of perfect nesting appear to be more focused. Previously hot-spot lines due to $\boldsymbol{Q}_0$-nesting have been predicted, but these are curved, not straight \cite{Elgazzar09, Oppeneer11}, therefore in this high-symmetry plane only a part of the hot-spot lines is clearly discernible. 
As we probe here just one of the $(1,1,0)$ directions \cite{note}, eight such lines can be expected on the FS. 
The presence of the QP intensity in the nesting loci indicates a pairing of states over $\boldsymbol{Q}_0$ on the involved $\Gamma$- and Z-centered FS sheets. 

A particularly bright QP spot is visible just above the Z point, where DFT calculations do not predict any FS sheet. Tentatively, we attribute this spot to a well-defined QP at $\boldsymbol{k} \simeq$ (0,0,1). Possibly this is the QP coupling to the aforementioned nested FSs. Here an Ising-like spin excitation has been proposed \cite{Oppeneer11}. Apart from the areas of strongly enhanced QP density there are also regions with weak QP intensity which are located between FS sheets at 
$k_z$=3.2\,{\an} and 2.95\,{\an},  (black circles in Fig.\ 4f, also marked in Fig.\ 4c). The origin of this QP density is not understood, probably it is related to interaction over an incommensurate nesting vector, possibly connecting the two $\Gamma$-centered FSs. 

Lastly we mention that previous ARPES experiments with fixed photon energies (\textit{viz.}\ 7\,eV \cite{Yokoya12, Yokoya10}, 21.2\,eV \cite{Santander09}, 29.5\,eV \cite{Dakovski11}, and 760\,eV \cite{Kawasaki11}) have revealed different Fermi vectors. Some of these photon energies are indicated in Fig.\ 4b and listed in Table I. We  stress here that the differences are in fact not mutually exclusive - since they can be reconciled as a direct consequence of the 3D FS of {\ur}. 

To conclude, using normal emission ARPES with tunable photon energies we have investigated the near FS behavior in the HO phase of {\ur}. Our study provides clear evidence that (i) the $5f$ electrons are in an itinerant configuration, (ii) the FS has a strong 3D topology, showing good agreement with DFT calculations.
In the HO state the FS appears folded over $\boldsymbol{Q}_0$=(0,0,1) and a substantial amount of intensity  is removed at $E_F$ along the (1,1,0) directions in the $k_z$=0 plane. (iii) The 3D nature of the FS supports various scattering vectors that have been measured with different photon energies. 
(iv) Enhanced QP intensities indicate essential pairing of FS states, particularly occurring at hot-spot lines. Our central conclusion is that HO is characterized by a strongly momentum dependent FS reconstruction.The FS removal and associated QP states appear as a signature of pairing of FS states on different parts of the Fermi surface that are joined by nesting vectors.

We thank A.F.\ Santander-Syro, F.\ Boariu, C.\ Bareille, A.V.\ Balatsky and Y.\ Dubi for helpful discussions.  This work was performed at Los Alamos National Laboratory under the auspices of the U.S. Department of Energy, Office of Basic Energy Sciences, Division of Materials Sciences and Engineering and LANL LDRD  Programs. P.M.O.\ was supported through the Swedish Research Council (VR) and P.R.\ was supported by the US DOE BES award DEFG02-84ER45872.

{}

\begin{thebibliography}{}
\bibitem{note} Note that the (1,1,0) $\Gamma$-M direction in the st BZ corresponds to $\Gamma$-X in the bct BZ.
\end{thebibliography}
\end{document}